 \documentclass[final,5p,times,twocolumn,numbers,square]{elsarticle}

\usepackage{lineno,hyperref}
\usepackage{amssymb,dsfont}
\usepackage{amsmath}
\usepackage{cancel}
\usepackage{cuted}
\usepackage{flushend}
\usepackage{graphicx,xcolor}
\usepackage[most]{tcolorbox}
\usepackage{array}
\usepackage{makecell}
\usepackage{multirow}
\usepackage{setspace}
\usepackage{colortbl}
\biboptions{sort&compress}

\newcommand{\tr}{\mathrm{tr}}
\newcommand{\tH}{\mathrm{H}}
\newcommand{\tvar}{\mathrm{var}}
\newcommand{\cP}{\mathcal{P}}
\newcommand{\Wg}{\mathrm{Wg}}

\makeatletter
\def\ps@pprintTitle{%
  \let\@oddhead\@empty
  \let\@evenhead\@empty
  \let\@oddfoot\@empty
  \let\@evenfoot\@oddfoot
}
\makeatother

\begin{document}

\begin{frontmatter}



\title{Exact mean and variance of the squared Hellinger distance for random density matrices}



\author{Vinay Kumar\corref{}}
\author{Kaushik Vasan\corref{}}
\author{Santosh Kumar\corref{cor1}}
\ead{skumar.physics@gmail.com}

\cortext[cor1]{Corresponding author}

\address{Department of Physics, Shiv Nadar Institution of Eminence, Gautam Buddha Nagar, Uttar Pradesh -- 201314, India}

\begin{abstract}
The Hellinger distance between quantum states is a significant measure in quantum information theory, known for its Riemannian and monotonic properties. It is also easier to compute than the Bures distance, another measure that shares these properties. In this work, we derive the mean and variance of the Hellinger distance between pairs of density matrices, where one or both matrices are random. Along the way, we also obtain exact results for the mean affinity and mean square affinity. The first two cumulants of the Hellinger distance allow us to propose an approximation for the corresponding probability density function based on the gamma distribution. Our analytical results are corroborated through Monte Carlo simulations, showing excellent agreement.
\end{abstract}




\end{frontmatter}




\section{\label{Introduction}Introduction}

In the rapidly evolving field of quantum information theory, the ability to quantify the difference between quantum states is of paramount importance. In this context, distance measures play a crucial role in a wide array of applications, including quantum state discrimination, quantum error correction, and the analysis of quantum entanglement~\cite{NC2010,RSI2016,MMPZ2008,PPZ2016,Wilde2013}. These measures provide the mathematical foundation for understanding the subtle differences between quantum states, which is essential for the development and implementation of quantum technologies.

Among the various distance measures used in quantum information, the (quantum) Hellinger distance stands out due to its elegant mathematical properties and practical computational advantages~\cite{LZ2004,BZ2017,DLH2011,MM2015,JF2018,Walczak2019}. However, in comparison to other distance measures, such as the relative entropy, the trace distance, the Hilbert-Schmidt distance, and the Bures distance, the Hellinger distance has been studied comparatively less in the context of random quantum states. Notably, exact results for up to second order statistics of both squared Hilbert-Schmidt and Bures-Hall distances are known for the case of random density matrices~\cite{ZS2005,Kumar2020,LAK2021,LK2023,LK2024}, , but, to the best of our knowledge, not for the Hellinger distance. In quantum information theory, Hellinger distance is the simplest and most straightforward generalization of the widely used classical Hellinger distance and happens to be both Riemannian and monotone. Consequently, it possesses desirable information theoretic properties and serves a bona-fide candidate for quantum correlations and distinguishability of states~\cite{LZ2004,RSI2016,JF2018}. One of the key advantages of the Hellinger distance over similar measures, such as the Bures distance, lies in its relatively easier computation, which makes it a more practical choice in scenarios where computational resources are limited or where rapid calculations are necessary~\cite{LZ2004}. 

In the following sections, we delve into the calculation of the Hellinger distance between two density matrices, where one can be fixed and the other random or both can be independently random. The random density matrices considered here are distributed according to Hilbert-Schmidt or Bures-Hall probability measures, which constitute two important class of probability distribution over the set of finite-dimensional mixed random states~\cite{Wooters1990,Hall1998,ZPNC2011,ZS2001,SZ2004,OSZ2010,CN2016,SK2019,Wei2020,SK2021}.

Hellinger distance between two density matrices $\rho_1$ and $\rho_2$ is defined as~\cite{LZ2004},
\begin{align}\label{Hellinger}
d_\tH(\rho_1,\rho_2) =\sqrt{\tr(\sqrt{\rho_1}-\sqrt{\rho_2})^2}
=\sqrt{2-2A(\rho_1,\rho_2)},
\end{align}
where 
\begin{align}
A(\rho_1,\rho_2)=\tr(\sqrt{\rho_1}\sqrt{\rho_2})
\end{align} 
is referred to as the affinity~\cite{Kholevo1972}. Affinity also equals the square root of Holevo's ``just-as-good fidelity", $F_\tH(\rho_1,\rho_2)$~\cite{Wilde2018} , i.e. $A(\rho_1,\rho_2)=\sqrt{F_\tH(\rho_1,\rho_2)}$. Therefore, we may refer to $d_\tH(\rho_1,\rho_2)$ as the Hellinger-Holevo distance. It assumes the minimum value of 0 when the two
states $\rho_1$ and $\rho_2$ are identical and the maximum value of $\sqrt{2}$ when the two states are supported on orthogonal subspaces. In some references, a factor of $1/\sqrt{2}$ is included in the definition given above in Eq.~\eqref{Hellinger}, which makes if vary from 0 to 1. In the following, we will deduce the statistics of the squared distance, viz., $D_\tH=d_H^2$.

\begin{figure*}[!ht]
\centering
\includegraphics[width=0.9\linewidth]{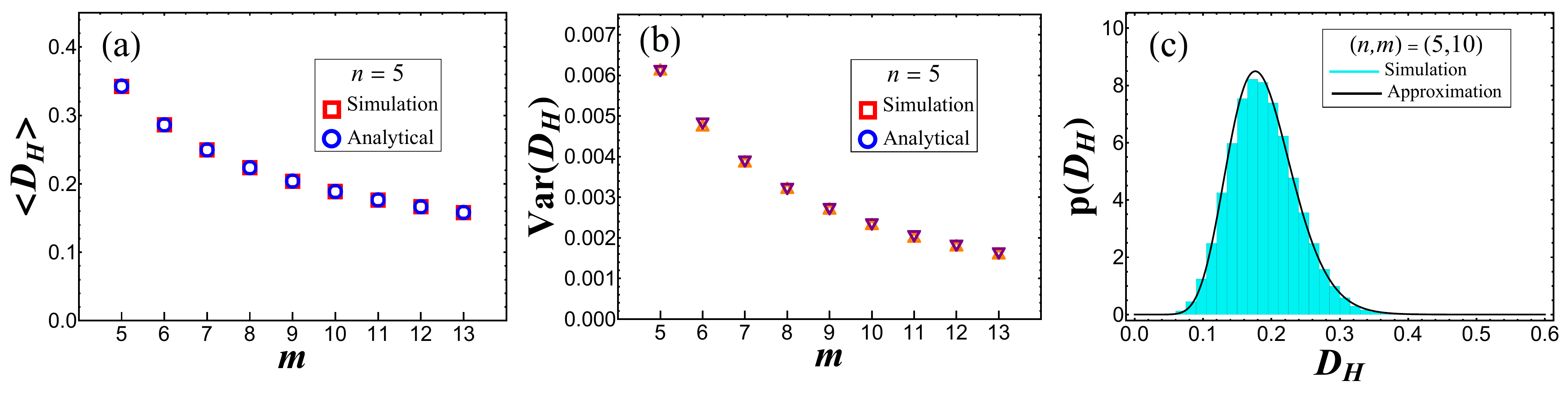}
\caption{Statistics of the squared Hellinger distance $D_\tH$ between a fixed density matrix and a Hilbert-Schmidt distributed random density matrix, both of dimension $n=5$. The fixed density matrix has eigenvalues $(7/100,16/100,17/100,23/100,37/100)$. Panels (a) and (b) show a comparison between exact analytical results and simulations for the mean and variance of $D_\tH$ as a function of ancilla dimension $m$ of the random density matrix. Panel (c) contrasts the distribution of $D_\tH$ obtained from simulation, for $(n,m)=(5,10)$, with the analytical approximation based on gamma distribution.}
\label{HSF}
\end{figure*}

\section{\label{SecRDM}Random Density Matrices}

The notion of random states is an important one in the areas of quantum chaos and quantum information theory~\cite{Wooters1990,Hall1998,ZPNC2011,ZS2001,SZ2004,OSZ2010,CN2016,SK2019,Wei2020,SK2021,Haake2010,BL2002}. In the case of finite-dimensional Hilbert-space set up, the random pure states are naturally characterized by the Haar measure on unitary group. However, for the mixed density matrix there is no such unique measure. The two prominent probability measures on this context are the Hilbert-Schmidt and Bures-Hall~\cite{Hall1998,ZS2001,ZPNC2011}. Both have been studied reasonably well and consequently several important results are know concerning them, including eigenvalues statistics. The density matrices distributed according to these measures can be obtained within the bipartite formalism, where one considers a composite system comprising two subsystems, say $S_1$ and $S_2$ (ancilla), described by $n$ and $m$-dimensional Hilbert spaces, respectively, with $n\le m$. The $n$-dimensional reduced density matrix obtained by partial tracing the ancilla $S_2$ on a random pure state of the composite system, belongs to the Hilbert-Schmidt ensemble~\cite{Hall1998,ZS2001,ZPNC2011,OSZ2010}. 
On the other hand, by applying operation of partial transpose on the superposition of a random pure state and its locally transformed copy gives rise to the Bures-Hall ensemble~\cite{Hall1998,ZPNC2011,OSZ2010,SK2019}. 

For obtaining statistical quantities of interest for the Hellinger distance, we will require the first two moments of the sum of square root of eigenvalues, $\sum_{i=1}^n\lambda_i^{1/2}=\tr\sqrt{\rho}$, of the random density matrices distributed according to Hilbert-Schmidt as well as Bures-Hall probability measures. Fortunately, these results are already known for both Hilbert-Schmidt ensemble~\cite{LK2023} and Bures-Hall ensemble~\cite{YHOW2024}. We compile them below for completeness.

\subsection{Hilbert-Schmidt Ensemble}\label{SecHSens}

The Hilbert-Schmidt ensemble is described by the probability density function (PDF)~\cite{ZS2001,SZ2004},
\begin{equation}
\label{HSdm}
    \mathcal{P}_\mathrm{HS}(\rho)\propto \delta(\tr\, \rho -1)\left( \det \rho\right)^{\alpha}\,\Theta(\rho),
\end{equation}
where 
\begin{equation}
\alpha=m-n
\end{equation} 
is the ``rectangularity" parameter, $\delta(\cdot)$ represents the Dirac delta function, and $\Theta(\cdot)$ represents the Heaviside theta function. The delta function ensures the unit-trace condition for $\rho$ and the theta function enforces the positive semi-definiteness condition on it. The Hilbert-Schmidt distributed density matrices can be constructed as~\cite{ZS2001,SZ2004,OSZ2010,ZPNC2011},
\begin{align}
\label{HSmatrixmodel}
\rho=\frac{GG^\dag}{\tr(GG^\dag)},
\end{align}
where $G$ is an $n\times m$-dimensional complex Ginibre random matrix, i.e., it contains independent and identically distributed zero-mean complex Gaussian elements.

The eigenvalues ($\{\lambda\}$) of the above random density matrix $\rho$ are governed by the joint probability density function~\cite{ZS2001,Hall1998},
\begin{equation}
\label{HSevdens}
P_\mathrm{HS}(\{\lambda\})=C_\mathrm{HS}\,\delta\left(\sum_{j=1}^n\lambda_j -1\right)\prod_{1\leq i < j \leq n}(\lambda_i-\lambda_j)^2\, \prod_{l=1}^n \lambda_l^\alpha,
\end{equation}
where $C_\mathrm{HS}$ is the normalization factor given by
\begin{equation}\label{C_HS}
C_\mathrm{HS}=\Gamma (nm) \left(\pi^{n(n-1)/2} \prod_{j=1}^n\Gamma(m-j+1)  \right)^{-1}.
\end{equation}
As stated above, we require the first two moments of $\tr \sqrt{\rho}$. For the Hilbert-Schmidt distributed random density matrix, we have~\cite{LK2023},
\begin{equation}\label{HSav1}
\left\langle \tr \sqrt{\rho}\right\rangle_\mathrm{HS}=\left\langle\sum_{r=1}^n \lambda_r^{1/2}\right\rangle_\mathrm{HS} =\frac{2}{(mn)_{1/2}}\sum_{j=1}^n\binom{1/2}{j}\binom{1/2}{j-1}\frac{(m)_{3/2-j}}{(n+1)_{-j}},
\end{equation}
where $\binom{a}{b}$ represents the binomial coefficient and $(a)_b=\Gamma(a+b)/\Gamma(a)$ is the Pochhammer symbol. Next, we have~\cite{LK2023},
\begin{align}
\label{HSav2}
\nonumber
&\left\langle ( \tr \sqrt{\rho}\right)^2\rangle_\mathrm{HS}
=\left<\left(\sum_{r=1}^n \lambda_r^{1/2}\right)^2\right>_\mathrm{HS}\\
&=1+\frac{2\,\Gamma^2(\alpha+\frac{3}{2})}{nm}\sum_{0\le j < k \le n-1}(\xi_{jj}\xi_{kk}-\xi_{jk}\xi_{kj}),
\end{align}
where
\begin{align}
\nonumber
\xi_{jk}
&=\frac{j!\, \binom{\frac{1}{2}}{j}}{(j+\alpha)!\, \binom{\frac{1}{2}}{k}\,\Gamma(\alpha+\frac{3}{2})}\sum_{l=0}^j\binom{\frac{1}{2}}{j-l}\binom{\frac{1}{2}}{k-l}\frac{\Gamma(l+\alpha+\frac{3}{2})}{l!}\\
&=\frac{j!}{(j+\alpha)!}\binom{\frac{1}{2}}{j}^2\,_3F_2\left(\alpha+\frac{3}{2},-j,-k;\frac{3}{2}-j,\frac{3}{2}-k;1\right).
\end{align}

\begin{figure*}[!h]
\centering
\includegraphics[width=0.9\linewidth]{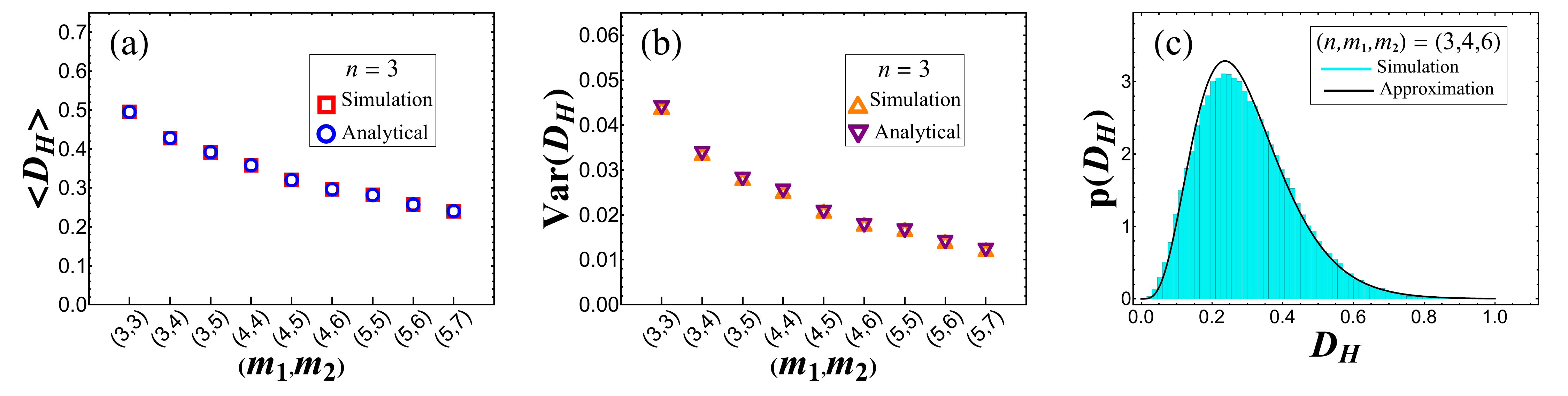}
\caption{Statistics of the squared Hellinger distance $D_\tH$ between two independent Hilbert-Schmidt distributed random density matrices of dimension $n=3$. In panels (a) and (b), compared between exact analytical and simulation results has been shown for the mean and variance of $D_\tH$ for varying ancilla dimensions $m_1,m_2$ of the two random density matrices. In panel (c) the distribution of $D_\tH$ obtained from simulation, for $(n,m_1,m_2)=(3,4,6)$, has been contrasted with the analytical gamma-distribution-based approximation.}
\label{HS}
\end{figure*}
\subsection{Bures-Hall Ensemble}\label{SecBHens}

The density matrices belonging to Bures-Hall ensemble are distributed according to the PDF~\cite{ZS2001,Hall1998},
\begin{equation}
\label{BHdm}
\mathcal{P}_\mathrm{BH}(\rho)\propto \delta(\tr\rho-1)(\det \rho)^\alpha \Theta(\rho)\int d[X]e^{-\tr \,\rho X^2},
\end{equation}
where $X$ is a random Hermitian matrix and $d[X]$ represents the product of differential of independent components in $X$. These density matrices can be generated as~\cite{OSZ2010,ZPNC2011,FK2016,SK2019},
\begin{align}
\label{BHmatrixmodel}
\rho=\frac{(\mathds{1}+U)GG^\dag(\mathds{1}+U^\dag)}{\tr[(\mathds{1}+U)GG^\dag(\mathds{1}+U^\dag)]},
\end{align}
where $G$ is a complex Ginibre random matrix as in the Hilbert-Schmidt case, and $U$ is a random unitary matrix from the probability measure proportional to $|\det(\mathds{1}+U)|^{2(m-n)}d\mu(U)$, with $d\mu(U)$ being the Haar-measure on the group of $n$-dimensional unitary matrices.

The eigenvalues of the Bures-Hall distributed density matrices adhere to the joint PDF given by~\cite{ZS2001,Hall1998}, 
\begin{equation}
\label{BHevdens}
 P(\{\lambda\}) = C_\mathrm{BH} \,\delta\left(\sum_i^n \lambda_i -1 \right)\prod_{1\leq i < j \leq n}\frac{(\lambda_i-\lambda_j)^2}{\lambda_i+\lambda_j}\prod_{k=1}^n\ \lambda_k^{\alpha-1/2},
\end{equation}
where $C_\mathrm{BH}$ is the normalization factor in this case, given by
\begin{equation}
C_\mathrm{BH} = \frac{2^{-n(n+2\alpha+1)}\pi^{n/2}}{\Gamma(n(n+2\alpha+2)/2)}\prod_{i=1}^n\frac{\Gamma(i+1)\Gamma(i+2\alpha+3)}{\Gamma(i+\alpha+1)}.
\end{equation}
In this case, the first moment of $\tr\sqrt{\rho}$ is known to be~\cite{YHOW2024},
\begin{align}\label{BHav1}
\nonumber
&\left\langle \tr \sqrt{\rho}\right\rangle_\mathrm{BH}=\left<\sum_{r=1}^n\lambda_r^{1/2}\right>_\mathrm{BH} \\
\nonumber
    &=\frac{1}{\pi\,(d)_{1/2}} \sum_{i=0}^{n-1} \frac{(i+1)_{1/2}\,(i+2\alpha+1)_{1/2}\,(i+\alpha+3/2)_{1/2}}{(n-i-1/2)_{1/2}\,(i+2\alpha+n+1)_{1/2}} \\
    &~~~~~~~~~~~~~~\times \left(1+\frac{i+\alpha+1/2}{i+\alpha+1} \right),
\end{align}
where $d$ is defined as,
\begin{equation}
    d = \frac{n(n+2\alpha)}{2}.
\end{equation}
The second moment of $\tr\sqrt{\rho}$ is given by~\cite{YHOW2024},
\begin{align}
\label{BHav2}
\nonumber
    &\left\langle (\tr \sqrt{\rho}\right)^2\rangle_\mathrm{BH}=\left<\left(\sum_{r=1}^n \lambda_r^{1/2}\right)^2\right>_\mathrm{BH} \\
    \nonumber
    &= 1+ \frac{1}{\pi^2 d}\sum_{i=0}^{n-1}\sum_{j=0}^{n-1}L_{i}L_{j} \Bigg[ \left(2+\frac{1}{2(i+\alpha+\frac{1}{2})}\right) \left(2+\frac{1}{2(j+\alpha+\frac{1}{2})}\right) \\
    \nonumber
    & - \frac{1}{2(i-j-\frac{1}{2})(j-i-\frac{1}{2})} \left(1+\frac{(i+\alpha+1)(j+\alpha+1)}{(i+\alpha+\frac{1}{2})(j+\alpha+\frac{1}{2})} \right) \\
&+\frac{j+\alpha+1}{(i+j+2\alpha+1)(i+j+2\alpha+2)(j+\alpha+\frac{1}{2})}\Bigg].
\end{align}
In the above equation, $L_{i}$ is defined as,
\begin{equation}
    L_{i} = \frac{(i+1)_{1/2}\,(i+\alpha+1/2)_{1/2}\,(i+2\alpha+1)_{1/2}}{(n+2\alpha+i+1)_{1/2}\,(n-i-1/2)_{1/2}}.
\end{equation}

\begin{figure*}[!ht]
\centering
\includegraphics[width=0.9\linewidth]{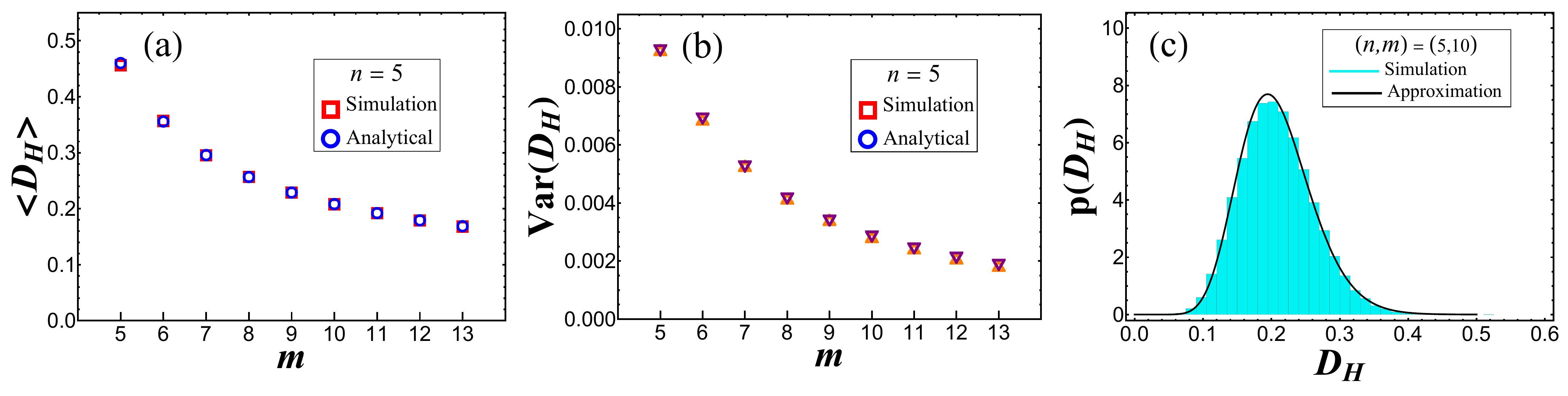}
\caption{Statistics of the squared Hellinger distance $D_\tH$ between a fixed density matrix and a Bures-Hall distributed random density matrix, both of dimension $n=5$. The considered fixed density matrix is same as in Fig.~\ref{HSF}. In panels (a) and (b), comparison between exact analytical and simulation results has been shown for the mean and variance of $D_\tH$ for varying ancilla dimension $m$ of the random density matrix. In panel (c) the distribution of $D_\tH$ obtained from simulation, for $(n,m)=(5,10)$, has been contrasted with the analytical gamma-distribution-based approximation.}
\label{BHF}
\end{figure*}
\section{\label{Stat_hell_dis} Statistics of squared Hellinger distance}

The average of the squared Hellinger distance can be obtained as,
\begin{align}\label{Distav1}
\nonumber
\langle D_\tH \rangle = 2-2\left<\text{tr}\left(\sqrt{\rho_1}\sqrt{\rho_2} \right)\right>\\
=2-2\langle A(\rho_1,\rho_2)\rangle,
\end{align}
and its variance turns out to be, 
\begin{align} \label{Distav2}
\nonumber
    \tvar (D_\tH)& = 4\big[\langle\left(\text{tr}\left(\sqrt{\rho_1}\sqrt{\rho_2} \right)\right)^2\rangle- \langle \text{tr}\left(\sqrt{\rho_1}\sqrt{\rho_2} \right)\rangle \big] \\
    &=4\big[\langle (A(\rho_1,\rho_2))^2\rangle - \langle A(\rho_1,\rho_2)\rangle \big].
\end{align}
Therefore, we need to calculate the mean affinity and mean square affinity in order to determine the average and variance of the squared Hellinger distance. Moreover, once we have obtained the above two quantities, we can approximate the PDF for $D_\tH$ using a gamma distribution by applying the cumulant matching method, viz.,
\begin{align}
p(x)=\frac{\eta^\zeta}{\Gamma(\zeta)}x^{\zeta-1}e^{-\eta x}.
\end{align}
Here $\zeta (>0)$ and $\eta (>0)$ are the shape and rate parameters, respectively, which can be obtained using the mean and variance as,
\begin{align}
\eta=\frac{\langle D_\tH \rangle}{ \tvar (D_\tH)},~~~\zeta=\frac{\langle D_\tH \rangle^2}{ \tvar (D_\tH)}.
\end{align}
The choice of the gamma distribution for approximation, instead of the common Gaussian, arises from the fact that $D_H$ is a non-negative quantity.

We now set on to obtain the required averages and begin by considering the eigen-decomposition of the random density matrix $\rho$,
\begin{equation}\label{rho_diagonalized}
    \rho = U\Lambda U^\dagger.
\end{equation}
Here, $\Lambda$ is diagonal matrix containing the eigenvalues $\{\lambda\}$ of $\rho$ and $U$ is a random unitary matrix. We also have, $\sqrt{\rho} = U\sqrt{\Lambda} U^\dagger$, with $\sqrt{\Lambda}$ being a diagonal matrix containing the positive square roots of $\{\lambda\}$, in view of $\rho$ being positive-semidefinite. Calculating the desired averages involves performing group integrals over the random unitary matrix. This is accomplished using Weingarten functions~\cite{Weingarten1978,Collins2003}, as detailed in the Appendices.

In the following, we separately examine the cases where one of the two density matrices is random and the other is fixed, as well as where both matrices are independent and random.

\begin{figure*}[!h]
\centering
\includegraphics[width=0.9\linewidth]{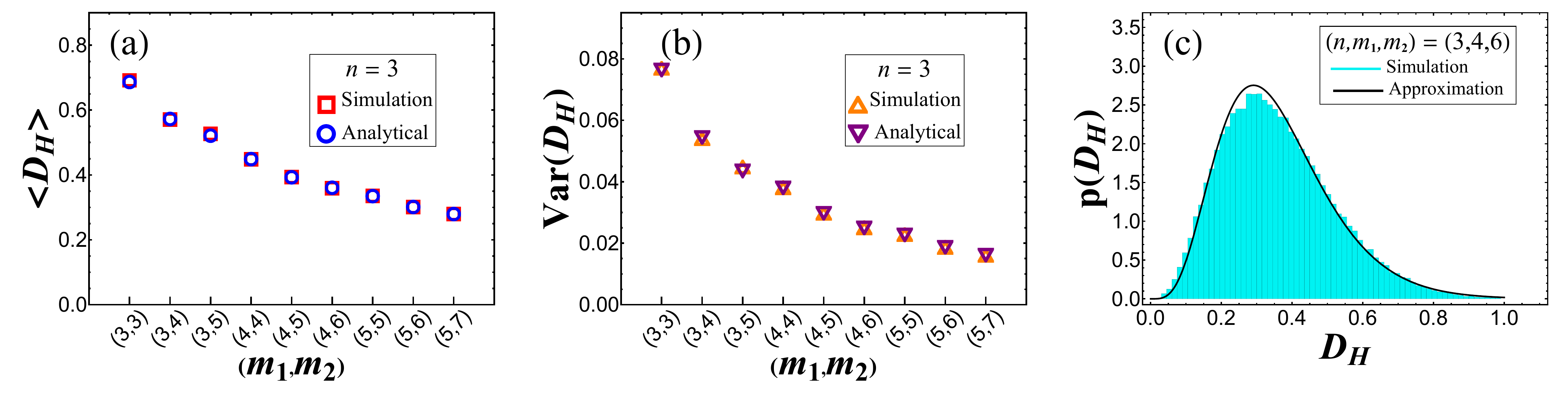}
\caption{Statistics of the squared Hellinger distance $D_\tH$ between two independent Bures-Hall distributed random density matrices of dimension $n=3$. In panels (a) and (b), compared between exact analytical and simulation results has been shown for the mean and variance of $D_\tH$ for varying ancilla dimensions $m_1,m_2$ of the two random density matrices. In panel (c) the distribution of $D_\tH$ obtained from simulation, for $(n,m_1,m_2)=(3,4,6)$, has been contrasted with the analytical gamma-distribution-based approximation.}
\label{BH}
\end{figure*}
\subsection{One random density matrix and one fixed density matrix}\label{sub:Statistics one random one fixed}

Let us consider a random density matrix $\rho$ belonging to either the Hilbert-Schmidt ensemble or the Bures-Hall ensemble, and a fixed density matrix $\sigma$. We first compute the mean affinity, i.e., $\langle A(\rho,\sigma)\rangle=\langle \tr(\sqrt{\rho}\sqrt{\sigma})\rangle$. We have,
\begin{equation} \label{Eq:average of fixed and random}
\langle A(\rho,\sigma)\rangle = \int d[\rho] \, \cP(\rho)\text{tr} \big(\sqrt{\rho}\sqrt{\sigma}\big),
\end{equation}
where $\cP(\rho)$ represents the PDF associated with either the Hilbert-Schmidt ensemble, Eq.~\eqref{HSdm}, or the Bures-Hall ensemble, Eq.~\eqref{BHdm}, and $d[\rho]$ denotes the product of differentials of the independent matrix elements of the Hermitian $\rho$, excluding the unit trace condition. The above integral can be recast in terms of integrals over the eigenvalues and the unitary matrices as,
\begin{equation}
\label{firstgroupint}
\langle A(\rho,\sigma)\rangle = \int d\{\lambda\}  P(\{\lambda\})\int d\mu(U) \tr\left(U\sqrt{\Lambda}\,U^\dag\sqrt{\sigma}\right).
\end{equation}
Here, $P(\{\lambda\})$ denotes the joint PDF of eigenvalues, described either by the Hilbert-Schmidt probability density, Eq.~\eqref{HSevdens} or the Bures-Hall probability density, Eq.~\eqref{BHevdens}. As described in~\ref{AppendixA}, the integral over the random unitary matrix can be performed in terms of Weingarten function $\Wg(s,n)$ to yield,
\begin{align}\label{meanAff1}
\nonumber
\langle A(\rho,\sigma)\rangle &= \Wg(1,n)\left(\tr(\sqrt{\sigma}) \right)\int P(\{\lambda\})d\{\lambda\}\sum_{i=1}^n \lambda_i^{1/2}  \\
    &=\frac{1}{n}\tr(\sqrt{\sigma})\left<\tr(\sqrt{\rho})\right>,
\end{align}
where we have used the result $\Wg(1,n)=1/n$. We can now use Eq.~\eqref{HSav1} or Eq.~\eqref{BHav1} for $\langle\tr\sqrt{\rho}\rangle$ to find the above average for Hilbert-Schmidt ensemble or Bures-Hall ensemble.

Next, we evaluate the expression for the mean square affinity, $\langle (A(\rho,\sigma))^2\rangle=\left\langle \left(\tr\big(\sqrt{\rho}\sqrt{\sigma}\big)\right)^2\right\rangle$. We have,
\begin{equation}
\langle (A(\rho,\sigma))^2\rangle = \int d[\rho]\cP(\rho){\left(\tr\big(\sqrt{\rho}\sqrt{\sigma}\big)\right)^2 }.
\end{equation}
Again, decomposing the above integral in terms of eigenvalues and unitary matrix $U$, we obtain
\begin{equation}
\label{secondgroupint}
\langle (A(\rho,\sigma))^2\rangle = \int d\{\lambda\}  P(\{\lambda\})\int d\mu(U) \left(\tr\left(U\sqrt{\Lambda}\,U^\dag\sqrt{\sigma}\right)\right)^2.
\end{equation}
The unitary group integral can again be performed in terms of the Weingarten functions~\cite{Weingarten1978,Collins2003}, as shown in \ref{AppendixB}, and we obtain,
\begin{equation} 
\label{meanAff2}
\begin{split}
    \langle (A(\rho,\sigma))^2\rangle &=  \int d\{\lambda\}  P(\{\lambda\}) \\
    &\times\Bigg[\Wg(1^2,n)\Bigg\{\Bigg(\sum_{i=1}^n \lambda^{1/2}_i \Bigg)^2\left(\tr\sqrt{\sigma}\right)^2 +1 \Bigg\} \\
    &+\Wg(2,n)\Bigg\{ \left(\tr\sqrt{\sigma}\right)^2 + \Bigg(\sum_{i=1}^n \lambda_i^{1/2} \Bigg)^2\Bigg\}\Bigg].
\end{split}
\end{equation}
This, upon using the values $\Wg(1^2,n)=1/(n^2-1)$ and $\Wg(2,n)=-1/[n(n^2-1)]$, and some rearrangement of terms, we arrive at,
\begin{align}\label{meanFid}
\nonumber
 \langle (A(\rho,\sigma))^2\rangle &= \frac{1}{n^2-1} \left(1-\frac{1}{n}(\tr\sqrt{\sigma})^2\right)\\
& +\frac{1}{n^2-1}\left((\tr\sqrt{\sigma})^2-\frac{1}{n}\right)\left\langle \left(\tr\sqrt{\rho}\right)^2\right\rangle.
\end{align}
The expression for the average involving the eigenvalues can now be used from Eq.~\eqref{HSav2} or Eq.~\eqref{BHav2} and hence we have the desired expression for the mean square affinity. 

\subsection{Two independent random density matrices}\label{statistics for two random}

We now determine the mean affinity and mean square affinity between two $n$-dimensional independent random density matrices $\rho_1$ and $\rho_2$. Both matrices can be drawn from either the Hilbert-Schmidt ensemble or the Bures-Hall ensemble, potentially with different ancilla dimensions (denoted $m_1$ and $m_2$). Alternatively, we may consider one density matrix from the Hilbert-Schmidt ensemble and the other from the Bures-Hall ensemble, allowing us to compare density matrices sampled from different distributions.

The mean affinity can be evaluated as
\begin{equation} \label{Eq:average of random and random final}
 \langle A(\rho_1,\rho_2)\rangle= \int \int d[\rho_1] d[\rho_2] \cP_1(\rho_1)\cP_2(\rho_2) \text{tr}\left(\sqrt{\rho_1}\sqrt{\rho_2}\right). 
\end{equation}
We perform integral over one of the density matrices first (say $\rho_2$), considering the other ($\rho_1$) to be fixed, which enables us to use Eq.~\eqref{meanAff1} from the previous section. Subsequently, the integral over the other density matrix can be carried out. As a result, we obtain overall,
\begin{align} \label{Eq:average of random and random final}
\nonumber
 \langle A(\rho_1,\rho_2)\rangle &= \int \int d[\rho_1] d[\rho_2] \cP_1(\rho_1)\cP_2(\rho_2) \text{tr}\left(\sqrt{\rho_1}\sqrt{\rho_2}\right) \\
    \nonumber
    &= \int d[\rho_1] P(\rho_1) \left(\frac{1}{n} \tr \sqrt{\rho_1}\, \langle \tr \sqrt{\rho_2}\rangle\right) \\
    &=\frac{1}{n}\left<\tr\sqrt{\rho_1}\right>\left<\tr \sqrt{\rho_2}\right>. 
\end{align}
We observe that the two averages factorize neatly, allowing us to substitute the expressions from Eq.~\eqref{HSav2} or Eq.~\eqref{BHav2} to evaluate them.

Next, we adopt the same strategy to evaluate the mean square affinity between two independent random density matrices $\rho_1$ and $\rho_2$. We have,
\begin{align} 
\nonumber
& \langle (A(\rho_1,\rho_2))^2\rangle = \int \int d[\rho_1] d[\rho_2] \cP_1(\rho_1)\cP_2(\rho_2) (\tr\left(\sqrt{\rho_1}\sqrt{\rho_2})\right)^2\\
\nonumber
&= \int  d[\rho_1] \cP_1(\rho_1)\frac{1}{n^2-1}\Bigg[   1-\frac{(\tr\rho_1)^2}{n}
+\left( (\tr\sqrt{\rho_1})^2-\frac{1}{n}  \right)\langle (\tr\sqrt{\rho_2})^2 \rangle    \Bigg] \\
&=\frac{1}{n^2-1}\Bigg[ \left<(\tr\sqrt{\rho_1})^2\right>\left<(\tr \sqrt{\rho_2})^2\right> -\frac{ \left<(\tr\sqrt{\rho_1})^2\right>}{n}-\frac{ \left<(\tr\sqrt{\rho_2})^2\right>}{n}  +1\Bigg],
\end{align}
where we used Eq.~\eqref{meanFid} in the second line to perform the $\rho_2$-integral. The expressions for the averages obtained in the previous section can now be applied to calculate the mean square affinity. The results for the mean affinity and mean square affinity can then be substituted into Eqs.~\eqref{Distav1} and~\eqref{Distav2} to obtain the mean and variance of the squared Hellinger distance. At this point, we should mention that Ref.~\cite{KNR2021} provides the large-$n$ asymptotic expression for the mean square affinity between two independent random density matrices, sampled from the Hilbert-Schmidt distribution and having identical ancilla dimension, as $[_2F_1(1/2,-1/2;2;n/m)]^4$. Our result agrees very well with it when evaluated for $n\gtrsim 10$.

\begin{figure*}
\centering
\includegraphics[width=0.9\linewidth]{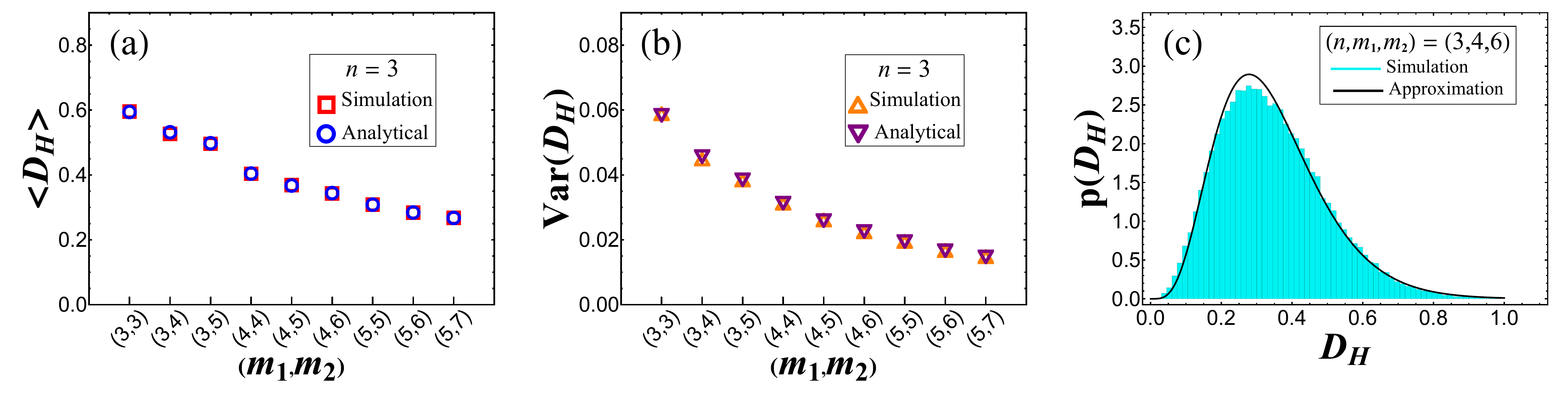}
\caption{Statistics of the squared Hellinger distance $D_\tH$ between two independent random density matrices of dimension $n=3$, one taken from the Hilbert-Schmidt distribution and the other from the Bures-Hall distribution. In panels (a) and (b), compared between exact analytical and simulation results has been shown for the mean and variance of $D_\tH$ for varying ancilla dimensions $m_1,m_2$ of the two random density matrices. In panel (c) the distribution of $D_\tH$ obtained from simulation, for $(n,m_1,m_2)=(3,4,6)$, has been contrasted with the analytical gamma-distribution-based approximation.}
\label{HSBH}
\end{figure*}
\section{\label{Res_&_dis}Comparison with matrix model simulations}

In this section, we compare our analytical results from the preceding sections with Monte Carlo simulations. Density matrices from the Hilbert-Schmidt ensemble can be generated numerically using the matrix model described in Eq.~\eqref{HSmatrixmodel}. For sampling density matrices from the Bures-Hall ensemble, one might consider using Eq.~\eqref{BHmatrixmodel}. However, the random unitary matrix involved in that equation does not follow the Haar measure in the general case ($m\ne n$), making it challenging to generate directly. Instead, we can generate the eigenvalues of density matrices from the Bures-Hall ensemble using the joint PDF given in Eq.~\eqref{BHevdens}, aided by a log-gas formalism-based Monte Carlo simulation. The corresponding random unitary matrix can then be obtained by conjugating a diagonal matrix containing these eigenvalues with a Haar-distributed random unitary matrix.

In Fig.~\ref{HSF}, we present the results for the squared Hellinger distance, $D_H$, where one of the two density matrices is fixed, and the other is randomly drawn from the Hilbert-Schmidt distribution. Panels (a) and (b) compare the analytical predictions for the mean and variance of $D_H$ with simulations for $n=5$ and various values of $m$, the ancilla dimension associated with the random density matrix. In panel (c), the gamma distribution-based approximation for the probability density of $D_H$ is shown alongside the simulation data for $(n,m)=(5,10)$.

Next, in Fig.~\ref{HS}, we analyze $D_H$ for the case of two independent random density matrices, both sampled from the Hilbert-Schmidt distribution. The results for the mean, variance, and probability density of $D_H$ are displayed. In panels (a) and (b), we consider $n=3$ and different combinations of $(m_1,m_2)$, corresponding to the ancilla dimensions of the two random density matrices. The simulation results in these panels are compared with the analytical predictions from the preceding sections. In panel (c), the probability density of $D_H$ is depicted for $(n,m_1,m_2)=(3,4,6)$ using matrix model simulations, and compared with the gamma distribution approximation.

Figure~\ref{BHF} illustrates the statistics of the squared Hellinger distance when one density matrix is fixed and the other is sampled from the Bures-Hall distribution. The panels show the same quantities as in Fig.~\ref{HSF}. Further, similarly to Fig.~\ref{HS}, in Fig.~\ref{BH}, we present the results for $D_\tH$ when both the density matrices are random, but now independently sampled from the Bures-Hall distribution. Finally, in Fig.~\ref{HSBH}, we examine the statistics of $D_\tH$ for two independent density matrices, with one sampled from the Hilbert-Schmidt distribution and the other from the Bures-Hall distribution.

In all these figures, we find very good agreement between the analytical predictions and simulation results for the mean and variance of $D_\tH$. Moreover, the gamma distribution based approximation for the distribution of $D_\tH$ works satisfactorily.

For the case of two independent density matrices, it is interesting to see that for identical values of parameters, the mean $D_\tH$ is the least when both are sampled from Hilbert-Schmidt distribution, and highest when both are sampled from Bures-Hall distribution. An intermediate value is observed for the case when one is sampled from the Hilbert-Schmidt distribution and the second from the Bures-Hall distribution. 

\section{\label{Sum&col}Summary and conclusion}

In this work, we investigated the squared Hellinger distance $D_\tH$ between pairs of density matrices. We derived exact analytical expressions for the mean and variance of $D_\tH$ under two distinct scenarios: when one density matrix is fixed while the other is random, and when both density matrices are random. The random density matrices are sampled from either the Hilbert-Schmidt or Bures-Hall distributions. Our derivation leverages Weingarten functions to perform the necessary unitary group integrals, supported by existing results for eigenvalue moments. Our analytical predictions are validated through Monte Carlo simulations across various parameter choices, demonstrating excellent agreement. Additionally, the gamma-distribution approximation for the probability density of $D_\tH$ shows satisfactory performance across all cases considered.

We believe that our results will be of significant interest to researchers in random matrix theory and quantum information, with potential applications in practical quantum information problems.

\section*{Acknowledgements}
V.K. thanks Shiv Nadar Institution of Eminence and SERB, DST, Government of India, for financial support. S.K. acknowledges support provided by SERB, DST, Government of India, via Project No. CRG/2022/001751.

\appendix

\section{Evaluation of the group integral in Eq.~\eqref{firstgroupint} } 
\label{AppendixA}

We focus on the group integral factor in Eq.~\eqref{firstgroupint}. We have,
\begin{align}
\nonumber
&\int d\mu(U) \tr\left(U\sqrt{\Lambda}\,U^\dag\sqrt{\sigma}\right)\\
\nonumber
&=\int d\mu(U) \sum_{i,j,k,l} U_{ij}(\sqrt{\Lambda})_{jk}(U^\dag)_{kl}(\sqrt{\sigma})_{li}\\
\nonumber
&=\int d\mu(U) \sum_{i,j,k,l} U_{ij}\, \lambda_j^{1/2}\delta_{jk}U^*_{lk}(\sqrt{\sigma})_{li}\\
&= \sum_{i,j,l} \lambda_j^{1/2}(\sqrt{\sigma})_{li}\int d\mu(U)U_{ij} U^*_{lj}\,.
\end{align}
We now use the following result to evaluate the unitary group integral in terms of the Weingarten function~\cite{Collins2003},
\begin{equation}
\int d\mu(U)U_{ij} U^*_{kl}=\Wg(1,n)\,\delta_{ik}\delta_{jl},
\end{equation}
and arrive at,
\begin{align}
\nonumber
&\int d\mu(U) \tr\left(U\sqrt{\Lambda}\,U^\dag\sqrt{\sigma}\right)\\
\nonumber
&=\sum_{i,j,l} \lambda_j^{1/2}(\sqrt{\sigma})_{li}\Wg(1,n) \delta_{il}\\
\nonumber
&=\Wg(1,n) \sum_{i,j} \lambda_j^{1/2}(\sqrt{\sigma})_{ii}\\
&=\Wg(1,n) \,\tr(\sqrt{\sigma})\sum_{j} \lambda_j^{1/2}.
\end{align}
This leads us to Eq.~\eqref{meanAff1}.

\section{Evaluation of the group integral in Eq.~\eqref{secondgroupint}} 
\label{AppendixB}

We examine the unitary group integral part in Eq.~\eqref{secondgroupint}. We find,
\begin{align}
\nonumber
&\int d\mu(U) \left(\tr\left(U\sqrt{\Lambda}\,U^\dag\sqrt{\sigma}\right)\right)^2\\
\nonumber
&=\int d\mu(U) \left( \sum_{i,j,l}   U_{ij} \lambda_j^{1/2}  U^*_{lj} (\sqrt{\sigma})_{li} \right)^2\\
\nonumber
&=\int d\mu(U) \sum_{i,j,l}  U_{ij} \lambda_j^{1/2}  U^*_{lj} (\sqrt{\sigma})_{li}\sum_{q,r,s} U_{qr} \lambda_r^{1/2}  U^*_{sr} (\sqrt{\sigma})_{sq}\\
&= \sum_{i,j,l}\sum_{q,r,s}  \lambda_j^{1/2}\lambda_r^{1/2}(\sqrt{\sigma})_{li}(\sqrt{\sigma})_{sq}\int d\mu(U)U_{ij} U_{qr} U^*_{lj}U^*_{sr}\,.
\end{align}
We now make use of the following result to evaluate the group integral in terms of Weingarten functions:
\begin{align}
\nonumber
 &\int d\mu(U)U_{ij}U_{kl }{U}^*_{pq}U^*_{rs}\\
 \nonumber
 =&\operatorname {Wg} (1^{2},n)(\delta _{ip}\delta _{jq}\delta _{kr}\delta _{l s} +\delta _{ir}\delta _{js}\delta _{kp}\delta _{l q})\\
& +\operatorname {Wg} (2,n)(\delta _{ip}\delta _{js}\delta _{kr}\delta _{l q}+\delta _{ir}\delta _{jq}\delta _{kp}\delta _{l s}).
\end{align}
This leads us to,
\begin{align}
\nonumber
&\int d\mu(U) \left(\tr\left(U\sqrt{\Lambda}\,U^\dag\sqrt{\sigma}\right)\right)^2\\
\nonumber
&=\int d\mu(U) \left( \sum_{i,j,l}   U_{ij} \lambda_j^{1/2}  U^*_{lj} (\sqrt{\sigma})_{li} \right)^2\\
\nonumber
&= \sum_{i,j,l}\sum_{q,r,s}  \lambda_j^{1/2}\lambda_r^{1/2}(\sqrt{\sigma})_{li}(\sqrt{\sigma})_{sq}\big[\operatorname{Wg}(1^2,n)(\delta_{il}\delta_{qs}+\delta_{is}\delta_{jr}\delta_{ql})\\
\nonumber
&~~~~~~~~~~~~~~ +\operatorname{Wg}(2,n)(\delta_{il}\delta_{jr}\delta_{qs}+\delta_{is}\delta_{ql})\big]\\
\nonumber
&=\operatorname{Wg}(1^2,n)\Bigg[\sum_{i,j,q,r}\lambda_j^{1/2}\lambda_r^{1/2}(\sqrt{\sigma})_{ii}(\sqrt{\sigma})_{qq}+\sum_{i,j,l}\lambda_j (\sqrt{\sigma})_{li}(\sqrt{\sigma})_{il}\Bigg]\\
\nonumber
&+\operatorname{Wg}(2,n)\Bigg[ \sum_{i,j,q} \lambda_j  (\sqrt{\sigma})_{ii} (\sqrt{\sigma})_{qq}+\sum_{i,j,l,r}\lambda_j^{1/2}\lambda_r^{1/2}(\sqrt{\sigma})_{li}(\sqrt{\sigma})_{il}\Bigg]\\
\nonumber
&=\operatorname{Wg}(1^2,n)\Bigg[\left(\sum_j\lambda_j^{1/2}\right)^2(\tr\sqrt{\sigma})^2  +\left(\sum_j\lambda_j\right) \Bigg]\\
&+\operatorname{Wg}(2,n)\Bigg[\left(\sum_j\lambda_j\right)  (\tr\sqrt{\sigma})^2+\left(\sum_j\lambda_j^{1/2}\right)^2 \tr\,\sigma \Bigg]
\end{align}
Now, since the trace of density matrices is unity, we have $\sum_j\lambda_j=1$ and also $\tr\,\sigma=1$, which leads us to Eq.~\eqref{meanAff2}.

\end{document}